\documentclass[twocolumn,letterpaper]{article}

\usepackage{iccv}
\usepackage{times}

\usepackage[utf8]{inputenc} 
\usepackage[T1]{fontenc}    
\usepackage{hyperref}       
\usepackage{url}            
\usepackage{booktabs}       
\usepackage{amsfonts}       
\usepackage{nicefrac}       
\usepackage{microtype}      
\usepackage{graphicx}
\usepackage{natbib}
\usepackage{doi}
\usepackage{algorithm}
\usepackage{algorithmic}
\usepackage{array}
\usepackage{multirow}
\usepackage{tabularx}
\usepackage{amsmath}
\usepackage{physics}
\usepackage{array,booktabs}

\iccvfinalcopy
\pdfoutput=1

\begin{document}
\title{CRASH: Raw Audio Score-based Generative Modeling for Controllable High-resolution Drum Sound Synthesis}

\author{Simon Rouard* \\
	Sony CSL - CentraleSupélec\\
    {\tt\small simon.rouard@student-cs.fr}
	\and
	Gaëtan Hadjeres* \\
	Sony CSL\\
	{\tt\small gaetan.hadjeres@sony.com}

}

\maketitle

{\let\thefootnote\relax\footnote{{*Equal contribution}}}

\begin{abstract}
In this paper, we propose a novel score-base generative model for unconditional raw audio synthesis.
Our proposal builds upon the latest developments on diffusion process modeling with stochastic differential equations, which already demonstrated promising results on image generation.
We motivate novel heuristics for the choice of the diffusion processes better suited for audio generation, and consider the use of a conditional U-Net to approximate the score function. While previous approaches on diffusion models on audio were mainly designed as speech vocoders in medium resolution, our method termed CRASH (Controllable Raw Audio Synthesis with High-resolution) allows us to generate short percussive sounds in 44.1kHz in a controllable way.
Through extensive experiments, we showcase on a drum sound generation task the numerous sampling schemes offered by our method (unconditional generation, deterministic generation, inpainting, interpolation, variations, class-conditional sampling) and propose the \emph{class-mixing} sampling, a novel way to generate “hybrid” sounds.
Our proposed method closes the gap with GAN-based methods on raw audio, while offering more flexible generation capabilities with lighter and easier-to-train models.

\end{abstract}
%
\section{Introduction and Related Work}\label{sec:introduction}
After multiple works in the spectral domain \cite{vasquez2019melnet,engel2019gansynth}, deep generative models in the waveform domain have recently shown the ability to produce high fidelity results with different methods: autoregressive \cite{oord2016wavenet,mehri2017samplernn}, flow-based \cite{prenger2018waveglow}, energy-based \cite{gritsenko2020spectral} or based on Generative Adversarial Networks \cite{donahue2019adversarial}.

In the task of generating drum sounds in the waveform domain, GAN-based approaches have been explored in \cite{donahue2019adversarial} and \cite{nistal2020drumgan}. However, the authors can generate only low-resolution 16kHz drum sounds which is often unacceptable for music producers. Interactive sound design is often a major motivation behind these works: in \cite{aouameur2019neural} the authors use Variational Autoencoders (VAE) in order to generate spectrograms of drums apply a principal component analysis on the latent space of the VAE in order to explore the drum timbre space. One of the disadvantages of this model is that the reconstruction of the sounds by the VAE tends to be blurry. In \cite{bazin2021spectrogram}, the authors use a VQ-VAE2 \cite{razavi2019generating} in order to perform inpainting on instrument sound spectrograms.

Score-based generative models \cite{vincent2011connection,ho2020denoising, song2019generative,song2021scorebased} propose a different approach to generative modeling, which consists in estimating the gradient of noise-corrupted data log-densities (score function): by iteratively denoising a sampled noise, these approaches obtained promising results, but mainly on image data. Moreover, the authors of Denoising Diffusion Implicit Model (DDIM) \cite{song2021ddim} use non-Markovian diffusion processes in order to accelerate the sampling of diffusion models.

To this day, only two score-based generative models in the waveform domain have been published \cite{kong2021diffwave, chen2020wavegrad} and they are mostly focused on the task of neural vocoding with conditioning on a mel-spectrogram. In \cite{kong2021diffwave}, the authors achieved the task of generating audio with an unconditioned model trained on the speech command dataset \cite{warden2018speech}. The inference scheme of \cite{kong2021diffwave} does not provide a flexible sampling scheme because it is trained on a fixed discrete noise schedule whereas \cite{chen2020wavegrad} is trained on a continuous scalar indicating the noise level.

In the image domain, \cite{song2021scorebased} generalizes the works of \cite{sohldickstein2015deep,ho2020denoising,song2019generative} by framing the noise corruption procedure as stochastic differential equation. 

Score-based generative models offer the following advantages over GAN-based approaches: 
\begin{itemize}
\item Training time is reduced and training is more stable since there is only one network to train.
\item Class-conditioning generation can be achieved by training a classifier a posteriori, which lets us train a model only one time.
\item Data can be mapped to a latent space without the need to train an additional encoder compared to GANs \cite{encoder2021tov}, which makes the interpolation between two given input data readily available with only one model. 
\end{itemize}
These properties alleviate us to search for directions in the latent space as in \cite{harkonen2020ganspace} or to directly hardcode conditional features in the architecture as in \cite{mirza2014conditional}. This easily controllable latent space permits sound design applications.
One downside of score-based models compared to GANs is their higher inference times to generate new samples. 

In this work, we extend the approach of \cite{song2021scorebased} and propose CRASH (Controllable Raw Audio Synthesis with High-resolution), a score-based generative model adapted to the waveform domain. On a drum sound dataset, the numerous capabilities offered by this architecture allows for musically-relevant sound design applications. Our contributions are the following: 
\begin{itemize}
\item A score-based model for unconditional generation that can achieve high fidelity 44.1 kHz drum sounds directly in the waveform domain,
\item The use of a noise-conditioned U-Net to estimate the score function,
\item A novel \emph{class-mixing} sampling scheme to generate "hybrid" sounds.
\item We provide a reparameterization of the SDE that shows that the DDIM deterministic sampling is another discretization of \cite{song2021scorebased} which leads to faster than real time sampling. 
\item Experimental and practical insights about the choice of the stochastic differential equation used to corrupt the data. 
\end{itemize}

\section{Background}
\label{sec:background}

\subsection{Score Based Modelling through Stochastic Differential Equations}
\begin{figure}[h]
    \centering
    \includegraphics[scale=0.2]{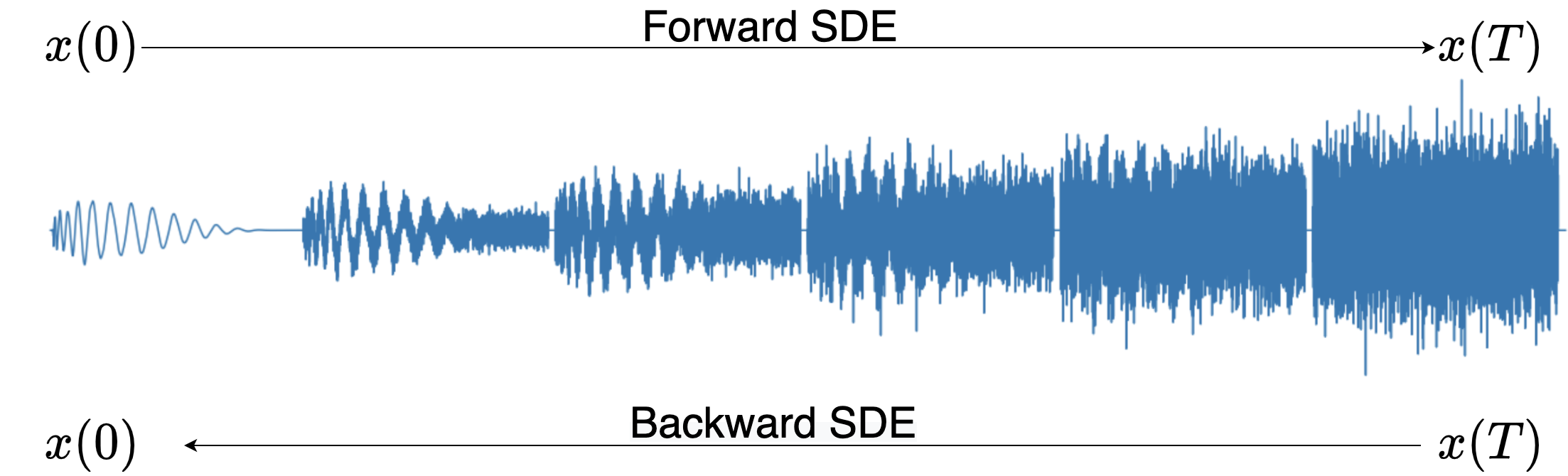}
    \caption{Illustration of the noising and denoising processes of a kick sound with a VP schedule}
    \label{fig:sde}
\end{figure}
\subsubsection{Forward Process}
Let $p_\text{data}$ be a data distribution. Diffusion models consist in progressively adding noise to the data distribution to transform it into a known distribution from which we can sample from as shown in  Fig.~\ref{fig:sde}. In \cite{song2021scorebased}, the authors formalize this noising process as the following \textbf{forward} Stochastic Differential Equation (SDE): 
\begin{equation}
\label{eq:sde1}
    \dd{\mathbf{x}} = f(t) \mathbf{x} \dd{t} + g(t) \dd{\mathbf{w}}
\end{equation} where $f(t)$ is a continuous negative function from $[0, T] \to \mathbb{R}^-$, $g(t)$ a continuous positive function from $[0, T] \to \mathbb{R}^+$, and $\mathbf{w}$ is a standard Wiener process. Such approach can be understood as a continuous-time generalization of Denoising Diffusion Probabilistic Models (DDPMs) \cite{sohldickstein2015deep,ho2020denoising} and denoising Score Matching with Langevin Dynamics (SMLD) \cite{song2019generative}.
For $\mathbf{x}(0) \sim p_\text{data}$, the transition kernel of Eq.~\ref{eq:sde1} is given by a normal distribution:

\begin{equation}
\label{eq:transition-kernel}
    p_t(\mathbf{x}(t) \mid \mathbf{x}(0)) = \mathcal{N}(\mathbf{x}(t); m(t)\mathbf{x}(0), \sigma^2(t)\mathbf{I}),
\end{equation}
where $m(t)$ and $\sigma(t)$ follow the system:
\begin{equation}
    \left\{
        \begin{array}{ll}
            \dv{m}{t} = f(t) m(t) \\
            \dv{\sigma^2(t)}{t} = 2f(t) \sigma^2(t) + g^2(t) 
        \end{array}
    \right.
\label{eq:system_m_sigma}
\end{equation}
with the following initial conditions $m(0) = 1$ and $\sigma^2(0) = 0$.

The solutions for $m(t)$ and $\sigma(t)$ are :
\begin{equation}
    \left\{
        \begin{array}{ll}
            m(t) = e^{\int_{0}^{t} f(s) \dd{s}} \\
            \sigma^2(t) = e^{\int_{0}^{t} 2f(s) \dd{s}} \int_{0}^{t} g^2(u) e^{\int_{0}^{u} -2f(s) \dd{s}} \dd{u}.
        \end{array}
    \right.
\label{solution_m_sigma}
\end{equation}

In \cite{song2021scorebased}, the authors define three types of SDEs which are presented in Tab.~\ref{sde_fg}.
\begin{table}[h!]
\begin{center}
\begin{tabular}{ | c | c | c | } 
  \hline
   & $f(t)$ & $g(t)$ \\ \hline
  VP & $-\frac{1}{2}\beta(t)$ & $\sqrt{\beta(t)}$ \\ \hline
  VE & $0$ & $\sqrt{\dv{[\sigma^2(t)]}{t}}$ \\ \hline
  sub-VP & $-\frac{1}{2}\beta(t)$ & $\sqrt{\beta(t)(1 - e^{-2\int_{0}^{t} \beta(s) \dd{s}})}$ \\ 
  \hline
\end{tabular}
\end{center}
\caption{Functions used in the VP, VE and sub-VP SDEs}
\label{sde_fg}
\end{table}

For the Variance Preserving (VP) and sub-Variance Preserving (sub-VP) schedules, $m(T)\approx 0$ and $\sigma(T) \approx 1$ which means that the original data distribution is transformed into a distribution close to a standard normal distribution i.e. $p_{T} \approx \mathcal{N}(\mathbf{0}, \mathbf{I})$. For the Variance Exploding (VE), $\sigma^2(T) \gg m \approx 1$ which means that the original data is not perceptible at $t=T$ and that $p_{T} \approx \mathcal{N}(\mathbf{0}, \sigma^2(T)\mathbf{I})$. 

\subsubsection{Generation with the Reverse Process}
In order to sample from the data distribution, we can sample $\mathbf{x}_T \sim p_T$ and apply the associated reverse time SDE \cite{anderson1982reverse} given by:
\begin{equation}
    \dd{\mathbf{x}} = [f(t)\mathbf{x} - g^2(t) \nabla_{\mathbf{x}} \log p_t(\mathbf{x})]\dd{t}+ g(t)\dd{\mathbf{\Tilde{w}}}
\label{eq:reverse_sde}
\end{equation} 
where $\mathbf{\Tilde{w}}$ is a standard Wiener process running backwards from T to 0 and $\dd{t}$ is an infinitesimal negative timestep. 

It means that by knowing $\nabla_{\mathbf{x}} \log p_t(\mathbf{x})$, we can use a discretization of Eq.~\ref{eq:reverse_sde} to sample $\mathbf{x}(0)$ from $p_0=p_\text{data}$. 

In practice, the score function $s(\mathbf{x}(t), \sigma(t)) = \nabla_{\mathbf{x}} \log p_{t}(\mathbf{x})$ is intractable and it is approximated by a neural network $s_\theta(\mathbf{x}(t), \sigma(t))$ parameterized by $\theta$. In order to train the network, \cite{vincent2011connection} shows that for any t, minimizing 
\begin{equation}
    \mathbb{E}_{p_{t}(\mathbf{x})} {\left\| s_\theta(\mathbf{x}, \sigma(t)) - \nabla_{\mathbf{x}} \log p_{t}(\mathbf{x}) \right\|}_2^2
\end{equation} is equivalent to minimizing 
\begin{equation}
\mathbb{E} {\left\| s_\theta(\mathbf{x}, \sigma(t)) - \nabla_{\mathbf{x}} \log p_{t}(\mathbf{x}(t) \mid \mathbf{x}(0)) \right\|}_2^2
\label{loss}
\end{equation}
where the expectation is over $\mathbf{x}(0) \sim p_\text{data}$, $\mathbf{x}(t) \sim p_{t}(\mathbf{x}(t) \mid \mathbf{x}(0))$,
and the latter distribution is given by Eq.~\ref{eq:transition-kernel}. 

Now, in order to train the network for all $t \in [0, T]$ we consider the following mixture of Eq.~\ref{loss} losses over all noise levels:
\begin{equation}
L(\theta)=\mathbb{E} \lambda(t) {\left\| s_\theta(\mathbf{x}(t), \sigma(t)) - \nabla_{\mathbf{x}(t)} \log p_{t}(\mathbf{x}(t) \mid \mathbf{x}(0)) \right\|}_2^2
\end{equation}
where we sample $t\sim[0,T]$, $\mathbf{x}(0) \sim p_\text{data}$, $\mathbf{x}(t) \sim p_{t}(\mathbf{x}(t) \mid \mathbf{x}(0))$ and where $\lambda(t)$ is a weighting function. 

In \cite{ho2020denoising, song2019generative, song2021scorebased}, $\lambda(t)$ is empirically set such that $\lambda(t)^{-1} \propto {\mathbb{E}{\left\| \nabla_{\mathbf{x}(t)} \log p_{t}(\mathbf{x}(t) \mid \mathbf{x}(0)) \right\|}^2_{2}} \propto \sigma^{2}(t)^{-1}$ while in \cite{durkan2021maximum} the authors show that the maximum likelihood estimator is obtained with $\lambda(t)=g^2(t)$ in $L(\theta)$. 

The training procedure is described in Alg.~\ref{alg:training_alg}, where we reparameterize our neural network as $\mathbf{\epsilon}_\theta(\mathbf{x}(t), \sigma(t)) := - \sigma(t)s_\theta(\mathbf{x}(t), \sigma(t))$ in order to estimate $\mathbf{\epsilon}$. 

\begin{algorithm}
\caption{Training procedure}
\label{alg:training_alg}
\begin{algorithmic} 
\WHILE{Training} 
\STATE Sample $t \sim \mathcal{U}([0, T]), \mathbf{x}(0) \sim p_{\text{data}}, \mathbf{\epsilon} \sim \mathcal{N}(\mathbf{0}, \mathbf{I})$
\STATE Compute $\mathbf{x}(t) = m(t)\mathbf{x}(0) + \sigma(t)\mathbf{\epsilon}$ 
\STATE Gradient descent on $\nabla_{\theta} {\left\|  \frac{\sqrt{\lambda(t)}}{\sigma(t)} [\mathbf{\epsilon}_\theta(\mathbf{x}(t), \sigma(t))  - \mathbf{\epsilon}] \right\|}_2^2$
\ENDWHILE
\end{algorithmic}
\end{algorithm}

Once the network is trained, a N-step discretization of the \textbf{reverse time} SDE is done in order to unconditionally generate samples. This process is described in Alg.~\ref{alg:sde_sampling}, it is non-deterministic since we obtain various sounds by starting from the same sample $\mathbf{x}(T)$.

\begin{algorithm}
\caption{Sampling via SDE}
\label{alg:sde_sampling}
\begin{algorithmic} 
\STATE Choose $N$, sample $\mathbf{x}_N \sim \mathcal{N}(\mathbf{0}, \sigma^2(T)\mathbf{I})$
\FOR{$i = N-1, ..., 0$} 
\STATE $t_i = T\frac{i}{N}, f_i = f(t_i), g_i = g(t_i), \sigma_i = \sigma(t_i)$
\STATE $\mathbf{x}_i = (1 - \frac{f_{i+1}}{N}) \mathbf{x}_{i+1} - \frac{g^2_{i+1}}{N\sigma_{i+1}} \mathbf{\epsilon}_\theta
(\mathbf{x}_{i+1}, \sigma_{i+1})$
\IF{$i>0$}
\STATE Sample $\mathbf{z}_{i+1} \sim \mathcal{N}(\mathbf{0}, \mathbf{I})$
\STATE $\mathbf{x}_i = \mathbf{x}_i + \frac{g_{i+1}}{\sqrt{N}} \mathbf{z}_{i+1}$
\ENDIF
\ENDFOR
\end{algorithmic}
\end{algorithm}

\subsection{Deterministic Sampling via Score based Ordinary Differential Equation}
As mentioned in \cite{song2021scorebased}, for any SDE, there exists a corresponding deterministic process which satisfies an ordinary differential equation (ODE):
\begin{equation}
    \dd{\mathbf{x}} = [f(t)\mathbf{x} - \frac{1}{2} g^2(t) \nabla_{\mathbf{x}} \log p_t(\mathbf{x})]\dd{t}
\label{eq:ode}
\end{equation}
This defines a flow $\phi^t$ such that the marginal distributions $\phi^t_*(p_\text{data})$ are identical to the ones obtained by applying the SDE of Eq.~\ref{eq:sde1}.
This mapping is interesting because it provides a latent representation for each $\mathbf{x} \sim p_\text{data}$. 

The procedure of sampling via the N-step discretization of the ODE is described in Alg.~\ref{alg:ode_sampling}. Moreover, we also experimented sampling by using the \texttt{scipy.integrate.solve\_ivp} solver with the RK45 method.

\begin{minipage}{0.49\textwidth}
\begin{algorithm}[H]
\caption{Sampling via ODE}
\label{alg:ode_sampling}
\begin{algorithmic} 
\STATE Choose $N$, sample $\mathbf{x}_N \sim \mathcal{N}(\mathbf{0}, \sigma^2(T)\mathbf{I})$
\FOR{$i = N-1, ..., 0$} 
\STATE $t_i = T\frac{i}{N}, f_i = f(t_i), g_i = g(t_i), \sigma_i = \sigma(t_i)$
\STATE $\mathbf{x}_i = (1 - \frac{f_{i+1}}{N}) \mathbf{x}_{i+1} - \frac{g^2_{i+1}}{2N\sigma_{i+1}} \mathbf{\mathbf{\epsilon}}_\theta
(\mathbf{x}_{i+1}, \sigma_{i+1})$
\ENDFOR
\end{algorithmic}
\end{algorithm}
\end{minipage}

\subsection{Inpainting}
Let's imagine that we don't like the attack of a kick (or any other part of a sound), the method of inpainting permits us to regenerate the desired part. In order to do that, we apply a reverse-time SDE or ODE discretization to an isotropic Gaussian and fix the part that we want to keep (with the associated noise corruption) after each denoising timestep. As presented in section \ref{sec:experiments}, we obtain very diverse and coherent results.
\begin{minipage}{0.49\textwidth}
\begin{algorithm}[H]
\caption{Inpainting via ODE or SDE}
\label{alg:inpainting}
\begin{algorithmic} 
\STATE Choose $N$, $U$ an inpainting mask, $\mathbf{x}_\text{fixed}$ a fixed sound, sample $\mathbf{x}_N \sim \mathcal{N}(\mathbf{0}, \sigma^2(T)\mathbf{I})$
\FOR{$i = N-1, ..., 0$} 
\STATE $t_i = T\frac{i}{N}, f_i = f(t_i), g_i = g(t_i), \sigma_i = \sigma(t_i), m_i=m(t_i)$
\IF{ODE Sampling}
\STATE $\mathbf{x}_i = (1 - \frac{f_{i+1}}{N}) \mathbf{x}_{i+1} - \frac{g^2_{i+1}}{2N\sigma_{i+1}} \mathbf{\mathbf{\epsilon}}_\theta
(\mathbf{x}_{i+1}, \sigma_{i+1})$
\ENDIF
\IF{SDE Sampling}
\STATE $\mathbf{x}_i = (1 - \frac{f_{i+1}}{N}) \mathbf{x}_{i+1} - \frac{g^2_{i+1}}{N\sigma_{i+1}} \mathbf{\mathbf{\epsilon}}_\theta
(\mathbf{x}_{i+1}, \sigma_{i+1})$
\IF{$i>0$}
\STATE Sample $\mathbf{z}_{i+1} \sim \mathcal{N}(\mathbf{0}, \mathbf{I})$
\STATE $\mathbf{x}_i = \mathbf{x}_i + \frac{g_{i+1}}{\sqrt{N}} \mathbf{z}_{i+1}$
\ENDIF
\ENDIF
\STATE Sample $\mathbf{z} \sim \mathcal{N}(\mathbf{0}, \mathbf{I})$
\STATE $\mathbf{x}_i \odot U = m_i (\mathbf{x}_\text{fixed} \odot U) + \sigma_i (\mathbf{z} \odot U)$
\ENDFOR
\end{algorithmic}
\end{algorithm}
\end{minipage}

\subsection{Interpolations}
The flexibility of SDEs and ODEs allows to compute interpolations between sounds. In fact, there exists an infinity of latent spaces indexed by $t \in [0, T]$. We present here two types of interpolations: ODE interpolation in the latent space of isotropic Gaussians and SDE interpolation in any t-indexed latent space.

\begin{figure}[h!]
    \includegraphics[scale=0.25]{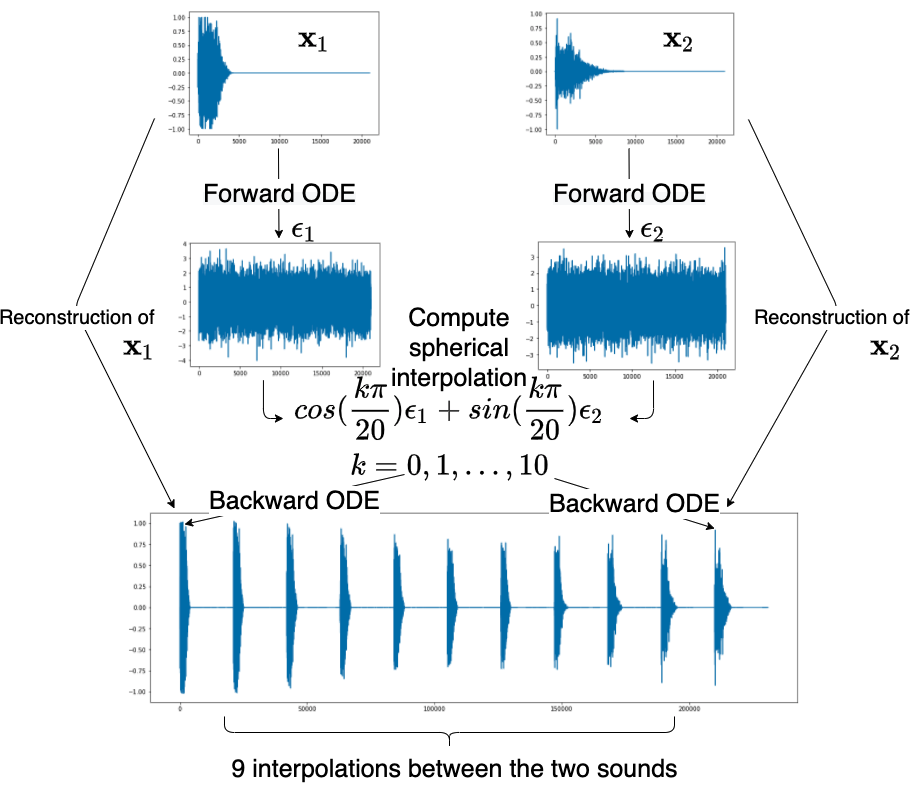}
    \caption{Interpolation of two sounds via Forward and Backward ODE}
    \label{fig:interpolation_schema}
\end{figure}

\subsubsection{ODE interpolation in the latent space of isotropic Gaussians}
Let $\mathbf{\epsilon}_1$ and $\mathbf{\epsilon}_2$ be two samples from a standard normal distribution of $\mathbb{R}^L$ where $L$ is our space dimension and $0\leq \lambda \leq1$. We consider the spherical interpolation $\mathbf{\epsilon}_\lambda = \lambda \mathbf{\epsilon}_1 + \sqrt{1-\lambda^2} \mathbf{\epsilon}_2$ and then apply the ODE sampling to it. We choose a spherical interpolation in order to preserve a variance close to 1 for $\mathbf{\epsilon}_\lambda$. 

Morever, if we want to interpolate two sounds $\mathbf{x}_1$ and $\mathbf{x}_2$, we can apply the Forward ODE in order to obtain the corresponding latent codes $\mathbf{\epsilon}_1$ and $\mathbf{\epsilon}_2$, apply the desired spherical interpolation and then apply an ODE sampling.

\subsubsection{ODE interpolation in a t-indexed latent space}
In \cite{kong2021diffwave}, the authors perform a linear interpolation between two sounds at a corresponding intermediate t-indexed latent space before applying Denoising Diffusion Probabilistic Model (DDPM is the discrete equivalent of a VP SDE). We adapt the method to the continuous framework with SDE and ODE. Here again, the interpolation can be done between two t-indexed latent codes or between sounds corrupted using the transition kernel of Eq.~\ref{eq:transition-kernel}.

\subsection{Class-Conditional sampling with a classifier}
For any class $y$, we can train a noise-conditioned classifier on corrupted data $\mathbf{x}(t)$. As a consequence, the output of the classifier gives us $p_t(y \mid \mathbf{x}(t))$ for each class $y$. We can use automatic-differentiation to differentiate this quantity and by the Bayes Formula, since $p(y)$ is constant for each class $y$, we have the following formula: 
\begin{equation}
\label{bayes_formula}
    \nabla_{\mathbf{x}} \log p_t(\mathbf{x} \mid y) = \nabla_{\mathbf{x}} \log p_t(\mathbf{x}) + \nabla_{\mathbf{x}} \log p_t(y \mid \mathbf{x})
\end{equation} 
As a consequence, we can generate samples of one class by solving this reverse time SDE: 
\begin{equation}
\label{class_cond_eq}
    \dd{\mathbf{\mathbf{x}}} = [f(t)\mathbf{x} - g^2(t)\nabla_{\mathbf{x}} \log p_t(\mathbf{x} \mid y)]\dd{t}+ g(t) \dd{\mathbf{\Tilde{w}}}
\end{equation}

This approach is flexible since it only requires to train a noise-conditioned classifier: there is no need to design and train a class-conditional score-based model as done in \cite{kong2021diffwave}.

\section{Reparameterizing the SDE and ODE and link between ODE and DDIM}

As shown in Sect.~\ref{sec:reparam}, for any SDE we can reparameterize it by using the associated perturbation kernel and obtain the following forward SDE for $\frac{\mathbf{x}}{m}$:
\begin{equation}
    \dd{\left(\frac{\mathbf{x}}{m}\right)} = \sqrt{\dv{t}(\frac{\sigma^2}{m^2})} \dd{\mathbf{w}}.
\label{eq:sde_snr}
\end{equation}

According to Eq.~\ref{eq:reverse_sde}, the associated reverse time SDE is (see Sect.~\ref{sec:reparam} for details): 
\begin{equation}
    \dd{\left(\frac{\mathbf{x}}{m}\right)}= 2 \dv{t}(\frac{\sigma}{m})\mathbf{\epsilon}(\mathbf{x}, \sigma) \dd{t}+ \sqrt{\dv{t}(\frac{\sigma^2}{m^2})} \dd{\mathbf{\Tilde{w}}}
\end{equation}
where $\mathbf{\epsilon}(\mathbf{x}, \sigma) := - \sigma(t)\nabla_{\mathbf{x}} \log p_t(\mathbf{x})$.

In the same way, if we compute the associated deterministic ODE, we obtain:
\begin{equation}
    \dd{\left(\frac{\mathbf{x}}{m}\right)}=\dd{\left(\frac{\sigma}{m}\right)}\mathbf{\epsilon}(\mathbf{x}, \sigma).
\label{eq:ode_reparam}
\end{equation}

Moreover, this ODE is a refactored version of Eq.~\ref{eq:ode} divided by $m$. It means that Eq.~\ref{eq:ode} and Eq.~\ref{eq:ode_reparam} encode the same latent representation. 
Now, by integrating Eq.~\ref{eq:ode_reparam} between $t_i$ and $t_{i+1}$ (and by writing $\mathbf{x}_{i} := \mathbf{x}(t_i)$, $m_i := m(t_i)$, $\sigma_i := \sigma(t_i)$), we obtain:
\begin{equation}
\label{eq:ddim_int}
\begin{aligned}
    \frac{\mathbf{x}_{i+1}}{m_{i+1}} - \frac{\mathbf{x}_{i}}{m_{i}} = \int_{t_i}^{t_{i+1}} \dv{t}\left(\frac{\sigma(t)}{m(t)}\right) \mathbf{\epsilon}(\mathbf{x(t)}, \sigma(t)) \dd{t}\\
    \approx \int_{t_i}^{t_{i+1}} \dv{t}\left(\frac{\sigma(t)}{m(t)}\right) \mathbf{\epsilon}(\mathbf{x}_{i+1}, \sigma_{i+1}) \dd{t}\\
    = (\frac{\sigma_{i+1}}{m_{i+1}} - \frac{\sigma_{i}}{m_{i}})\mathbf{\epsilon}(\mathbf{x}_{i+1}, \sigma_{i+1})
\end{aligned}
\end{equation}

This discretization is exactly the deterministic one from Denoising Diffusion Implicit Models \cite{song2021ddim} (DDIM). Empirically, Alg.~\ref{alg:ddim_sampling} gives great samples with only 20 or 30 steps which permits to sample faster than real time. This comes from the fact that the only approximation error is due to $\mathbf{\epsilon}(\mathbf{x}(t), \sigma(t)) \approx \mathbf{\epsilon}(\mathbf{x}_{i+1}, \sigma_{i+1})$ between $t_i$ and $t_{i+1}$.

\begin{algorithm}[H]
\caption{DDIM sampling}
\label{alg:ddim_sampling}
\begin{algorithmic} 
\STATE Choose $N$, sample $\mathbf{x}_N \sim \mathcal{N}(\mathbf{0}, \sigma^2(T)\mathbf{I})$
\FOR{$i = N-1, ..., 0$} 
\STATE $t_i = T\frac{i}{N}, m_i = m(t_i), \sigma_i = \sigma(t_i)$
\STATE $\mathbf{x}_i = \frac{m_i}{m_{i+1}} \mathbf{x}_{i+1} + (\sigma_i - \sigma_{i+1}\frac{m_i}{m_{i+1}}) \mathbf{\epsilon}_\theta
(\mathbf{x}_{i+1}, \sigma_{i+1})$
\ENDFOR
\end{algorithmic}
\end{algorithm}

\begin{algorithm}[H]
\caption{Reparameterized SDE Sampling}
\label{alg:ddim_sde_sampling}
\begin{algorithmic} 
\STATE Choose $N$, sample $\mathbf{x}_N \sim \mathcal{N}(\mathbf{0}, \sigma^2(T)\mathbf{I})$
\FOR{$i = N-1, ..., 0$} 
\STATE $t_i = T\frac{i}{N}, m_i = m(t_i), \sigma_i = \sigma(t_i)$
\STATE $\mathbf{x}_i = \frac{m_i}{m_{i+1}} \mathbf{x}_{i+1} + 2(\sigma_i - \sigma_{i+1}\frac{m_i}{m_{i+1}}) \mathbf{\epsilon}_\theta
(\mathbf{x}_{i+1}, \sigma_{i+1})$
\IF{$i>0$}
\STATE Sample $\mathbf{z}_{i+1} \sim \mathcal{N}(\mathbf{0}, \mathbf{I})$
\STATE $\mathbf{x}_i = \mathbf{x}_i + \sqrt{(\frac{\sigma_{i+1} m_i}{m_{i+1}})^2 - \sigma_{i}^2} \mathbf{z}_{i+1}$

\ENDIF
\ENDFOR
\end{algorithmic}
\end{algorithm}

\section{A discussion about Choosing the Right SDE: a generalization of the sub-VP SDE}
In this section $T=1$.
\label{sec:discussion}
\subsection{The Signal to Noise Ratio (SNR)}
As Eq.~\ref{eq:sde_snr} shows, any SDE for $\mathbf{x}$ is equivalent to a Variance Exploding SDE for $\frac{\mathbf{x}}{m}$. We define $\text{SNR}$ as the \emph{signal-to-noise ratio associated to the SDE} by $\text{SNR}(t) := \frac{m(t)^2}{\sigma(t)^2}$, as it completely determines the signal-to-noise ratio of $x(t)$ via $\text{SNR}(\mathbf{x}(t)) = \text{SNR}(t)  \mathbb{E}[\mathbf{x}(0)^2]$. The quantities defined by the variations of $\text{SNR}(t)$ are more interpretable than the functions $f$ and $g$ when working with the SDE and ODE reparameterizations of Eq.\ref{eq:sde_snr} and Eq.\ref{eq:ode_reparam}. We test different functions in the experiments. Moreover, once that the SNR is defined, we still need to provide a function $m(t)$ (or equivalently $\sigma(t)$). 

\subsection{About the relation relation between $m(t)$ and $\sigma(t)$}
The VP SDE is the continuous version of the Denoising Diffusion Probabilistic Model (DDPM) used in \cite{ho2020denoising} \cite{kong2021diffwave} \cite{chen2020wavegrad}. One of the main features of this model is that the mean coefficient $m(t)$ of the perturbation kernel is linked to the standard deviation $\sigma(t)$ (or noise-level) by the following equation $m(t)=\sqrt{1-\sigma^2(t)}$. 

Moreover, without mentioning this fact, in \cite{song2021scorebased} the authors introduce the sub-VP SDE which is characterized by the following formula $m(t) = \sqrt{1 - \sigma(t)}$. They obtained their best results with this schedule. This formula leads to a $\text{SNR}$ that decays faster and that has a higher limit value in $t=1$. Since $\beta$ is fixed, it also leads to a slowly increasing $\sigma$ function near $t=0$. (See Fig.\ref{fig:sigma})

In this work, we explore four relations between $m$ and $\sigma$ described in Tab.~\ref{sde_fg_gen} in order to study the influence of the decay of the ${\text{SNR}}$ ratio. We also write the functions $f(t)$ and $g(t)$ for each of these 4 relations. For the rest of the paper we take the convention $f(t):=-\frac{1}{2}\beta(t)$ in order to compare the VP and sub-VP schedules with ours.

\begin{table}[h!]
\begin{center}
\scalebox{0.7}{%
\begin{tabular}{ | c | c | c | } 
  \hline
   $m$-$\sigma$ relation & $f(t)$ & $g(t)$ \\ \hline
  $m=\sqrt{1-\sigma^2}$ (VP) & $-\frac{1}{2}\beta(t)$ & $\sqrt{\beta(t)}$ \\ \hline
  $m=\sqrt{1-\sigma}$ (sub-VP) & $-\frac{1}{2}\beta(t)$ & $\sqrt{\beta(t)(1 - e^{-2\int_{0}^{t} \beta(s) \dd{s}})}$ \\ \hline
    $m=1-\sigma$ (sub-VP 1-1)& $-\frac{1}{2}\beta(t)$ & $\sqrt{\beta(t)(1 - e^{-\frac{1}{2}\int_{0}^{t} \beta(s) \dd{s}})}$ \\ \hline
  $m=(1-\sigma)^2$ (sub-VP 1-2)& $-\frac{1}{2}\beta(t)$ & $\sqrt{\beta(t)(1 - \frac{3}{2}e^{-\frac{1}{2}\int_{0}^{t} \beta(s) \dd{s}}+\frac{1}{2}e^{-\int_{0}^{t} \beta(s) \dd{s}})}$ \\ \hline

\end{tabular}}
\end{center}
\caption{Functions used in the VP, sub-VP and generalized sub-VP SDEs. We give the general formulas in section \ref{sec:sub-vp_gen}.}
\label{sde_fg_gen}
\end{table}

\subsection{Choosing the right functions for the SDE}
\begin{figure}[h!]
    \centering
    \includegraphics[scale=0.32]{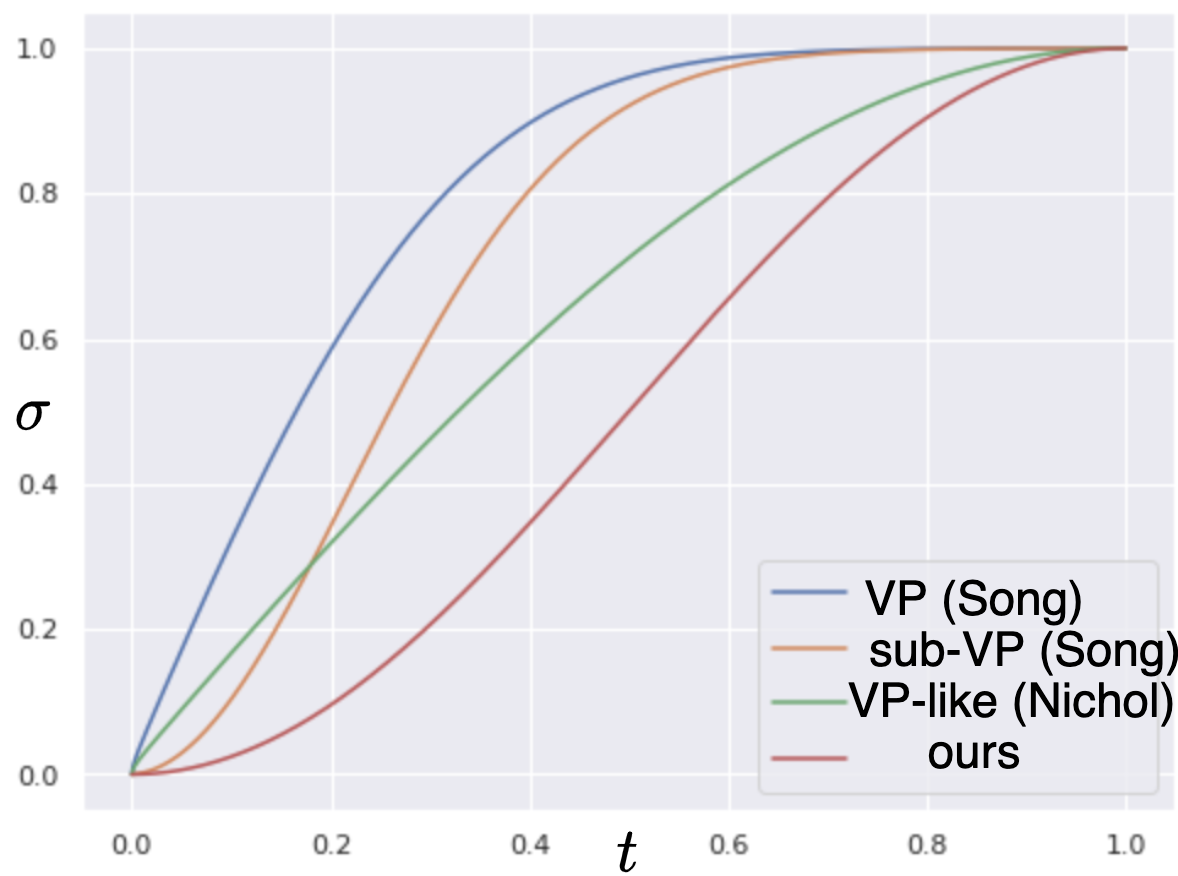}
    \caption{Different choices for the $\sigma(t)$ function}
    \label{fig:sigma}
\end{figure}

Choosing a particular relation between $m$ and $\sigma$, imposes a relation between $g$ and $\beta$. The remaining free parameter is the function $\beta$, needed to fully define the SDE.
In \cite{song2021scorebased}, the authors use a linear schedule for $\beta(t)$ because it is the continuous generalization of DDPMs. As presented in Fig.~\ref{fig:sigma}, this choice leads to a $\sigma(t)$ function that rapidly grows to its maximum. In \cite{nichol2021improved}, the authors mention this fast growing $\sigma$ function as a potential shortcoming and propose a smoother function (the green one in Fig.~\ref{fig:sigma}).

Our approach differs from \cite{song2021scorebased} in that the definition of our SDE is motivated by choosing a relatively smooth increasing function $\sigma(t)$ such as $\sigma(0) = 0$ and $\sigma(1) = 1 - \epsilon$ (where $\epsilon$ is a small constant), together with a $m$-$\sigma$ relation, from which all other quantities can be computed as shown in Tab.~\ref{sde_fg_gen}. 
If the two approaches are equivalent, we believe that these quantities are more interpretable. 
In the regime of a small number of discretization steps, a slow increasing function may induce less approximation errors.  
For our experiments we propose $\sigma(t)=\frac{1}{2}[1-\cos((1-s)\pi t)]$ with $s=0.006$ which is the red plot in Fig.~\ref{fig:sigma}. We also sample t in the interval $[\eta, 1]$ during the training where $\eta$ is chosen such that $\sigma(\eta) = 10^{-4}$ because $10^{-4}$ is imperceptible.  

\section{Class-mixing sampling with a classifier}

Drum classes are not perfectly distinct. For instance, the dataset contains drum sounds that are percussive enough to be seen as kicks but also sufficiently brilliant to be seen as snares and some kicks are combined with a hi-hat sound. We observe that our classifier (at the noise-level $\sigma=0$) sometimes outputs a mixed classes such as [0.3, 0.3, 0.4] and that it aligns well with our feeling when hearing the sound. 

We introduce the Class-Conditional sampling to a mixture of classes: 
For a given noisy sound $\mathbf{x}(t)$, the vector $\nabla_{\mathbf{x}(t)} \log p_t(y_i \mid \mathbf{x}(t))$ points out to the direction of the class $y_i$ in the noisy t-indexed latent space. Now, assuming that we have N classes $(y_i)_{i=1, …, N}$, let $(\lambda_i)_{i=1, …, N}$ be positive real numbers such as $\sum_{i=1}^N \lambda_i = 1$, we define a mixture of classes that we note $\{(y_i, \lambda_i)\}$ and the associated vector:
\begin{equation}
    \nabla_{\mathbf{x}} \log p_t(\{(y_i, \lambda_i)\} \mid \mathbf{x}) := \sum_{i=1}^N \lambda_i \nabla_{\mathbf{x}} \log p_t(y_i \mid \mathbf{x})
\end{equation}
In practice, we put this term in equation \ref{bayes_formula} in replacement of the last term and use equation \ref{class_cond_eq} to sample class-mixed audios. It gives us interesting results with a great palette of sounds. 

\section{Architecture}
\label{sec:architecture}
\subsection{Conditioned U-Net}
\label{subsec:unet}
Our model architecture is a conditioned U-Net \cite{MeseguerBrocal2019ConditionedUNetIA}, originally proposed for source separation. It takes two inputs: the noise level $\sigma(t)$ and the noisy audio $\mathbf{x}(t)$. The noise-level is encoded by Random Fourier Features \cite{tancik2020fourier} followed by a Multi-Layer Perceptron. The noisy audio goes into FiLM-conditioned \cite{perez2017film} Downsampling Blocks. Then, the signal goes into Upsampling Blocks that receive skip connections from the DBlocks of same levels. The output of the network is the estimated noise $\mathbf{\epsilon}_{\text{estimated}}$. 

\begin{figure}[h!]
    \includegraphics[scale=0.28]{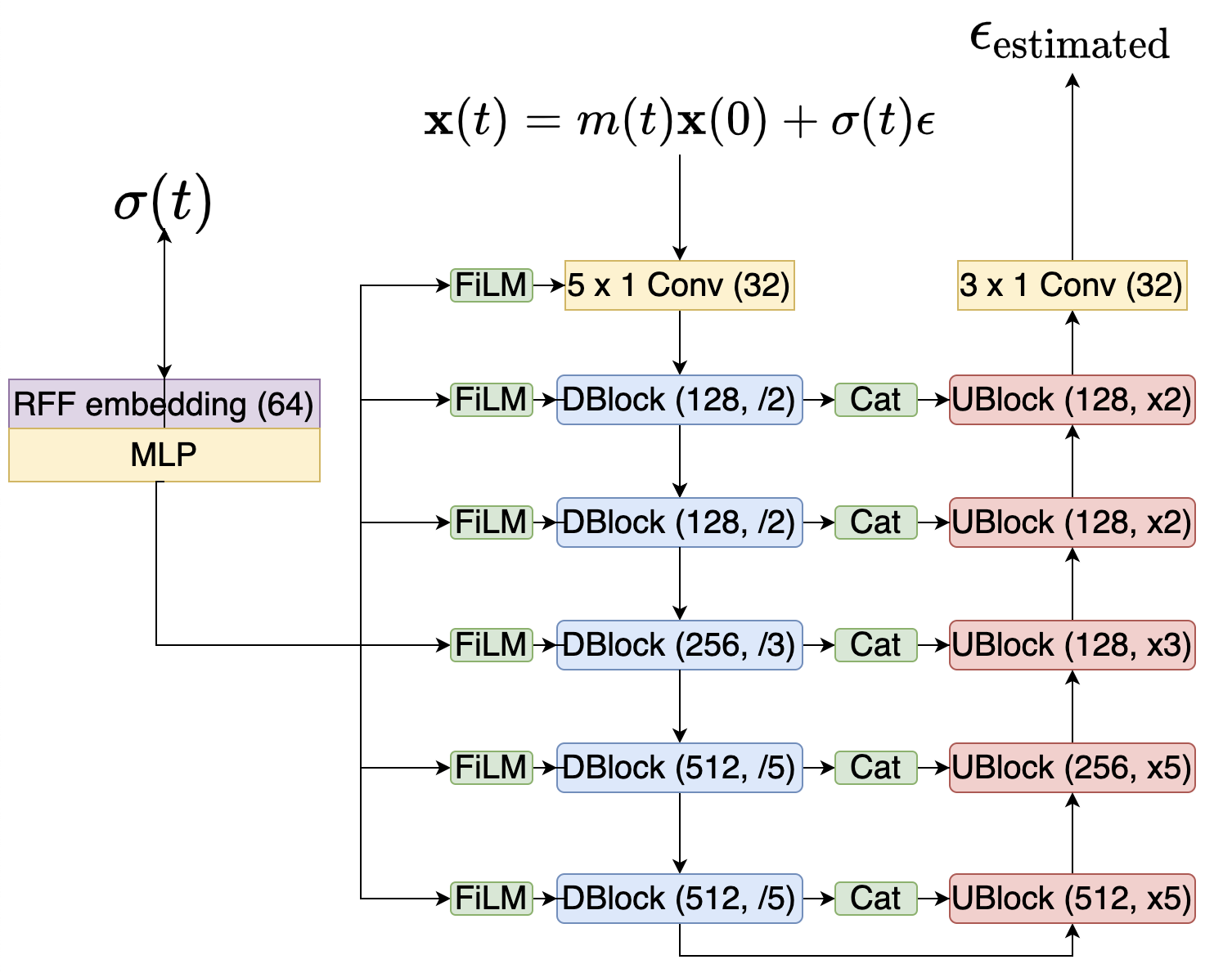}
    \caption{Architecture of the Conditioned U-Net}
    \label{fig:architecture}
\end{figure}

This bears similarities with the architecture from \cite{kong2021diffwave} which has a similar succession of blocks with dilated convolutions but no downsampling or upsampling layers, which makes it slow in terms of computation. The architecture from \cite{chen2020wavegrad} has a U-Net-like shape \cite{ronneberger2015unet}, but heavily depends on the spectrogram conditioning and relies on a different noise-conditioning scheme.
The $\sigma$-conditioned U-Net architecture seems to retain advantages from both approaches and is particularly suited for unconditional generation (see Fig.~\ref{fig:architecture}). 
The details of the architecture are presented in Sect.~\ref{sec:details}.

\subsection{Noise conditioned classifier}
Our noise-conditioned classifier closely mimics the architecture of our Conditioned U-Net presented in in Sect.~\ref{subsec:unet}.
The classifier is composed of a succession of FiLM-conditioned DBlocks followed by a projection layer and a softmax. Parameters for this architecture are presented in Sect.~\ref{sec:ncclassif}.

\section{Experiments and results}
\label{sec:experiments}
\subsection{Dataset}
For this work, we use an internal non-publicly available dataset of drum sounds which has also been used in \cite{nistal2020drumgan}. It is composed of approximately 300.000 one-shot kick, snare and cymbal sounds in equal proportions. The samples have a sample rate of 44.1kHz and are recorded in mono. We restricted and padded the audio to 21.000 time-steps because most sounds last less than 0.5 second. We used 90\% of the dataset in order to train our model. 

\subsection{Models and process}
We evaluate the influence of $\sigma(t)$ and four $m$-$\sigma$ schedules. The training of the network is done with a learning rate of $2.10^{-4}$ and the Adam optimizer. In parallel, smoothed weights with exponential moving average (EMA) with a rate of 0.999 are computed and saved at each step. For each model, the network is trained for about 120 epochs and the weights are saved each 8 epochs. We generated drum sounds with the regular weights and with the EMA weights and we observed the same phenomenon as in \cite{song2020improved}: for the regular weights the quality of the sounds is not necessarily increasing with the training time whereas the EMA weights provide better and more homogeneous Fréchet Audio Distance \cite{kilgour2019frechet} (FAD) during training\footnote{We use the original implementation \cite{kilgour2019frechet} available at \url{https://github.com/google-research/google-research/tree/master/frechet_audio_distance} }. 

After generating 2700 sounds for each checkpoint of each model, we choose the best checkpoints and generate 27000 drum sounds for each. It takes 12 hours to generate 27000 drum sounds on a Nvidia RTX-3090 GPU with an ODE or SDE schedule of 400 steps and batches of 180 sounds per generation (maximum memory capacity). In comparison, it takes around 5 hours with the scipy solver and it takes only 1.5 hours a DDIM 50 steps discretization which is faster than real time (27000 drum sounds represents 3.5 hours of audio).

\subsection{Quantitative Results}

We report the FAD (lower is better) between the 27000 generated drum sounds and the test set for each unconditional generation with SDE and ODE (with a discretization of 400 steps), a DDIM discretization with 50 steps and the \texttt{scipy.integrate.solve\_ivp} solver with \texttt{rtol = atol = $10^{-5}$, method='RK45'} parameters in Tab.~\ref{FADs}. The cos schedule refers to the function $\sigma(t)=\frac{1}{2}[1-\cos((1-s)\pi t)]$ (the red one in Fig.~\ref{fig:sigma}) and the exp schedule corresponds to the function $\sigma(t)=\sqrt{1 - e^{-0.1 t - 9.95 t^2}}$ used in \cite{song2021scorebased} (the blue one in Fig.~\ref{fig:sigma}). 
Because adding Gaussian noise with a factor $10^{-4}$ is almost non perceptible, we also decided to compute the FAD between the generated samples and a noisy version of the test set where we corrupted the sounds with a $10^{-4}$ level of Gaussian noise. The DDIM sampling and Scipy solver generate samples that are a bit noisy at a non-perceptible level, this is why they perform better on the noisy test set. Moreover, note that the FAD between the test set and the noisy test set is 0.72 which means that the FAD is a very sensitive metric. As a consequence, by looking at Tab.~\ref{FADs} we can only say that the "cos schedule" is more adapted than the "exp schedule" from \cite{song2021scorebased} to audio data for equally spaced discretization steps because there are more steps near $\sigma=0$ which is crucial in order to obtain non-noisy sounds. We cannot really say if some sub-VP schedules are better than others and we think that the comparison should be done on image data because the metrics are more meaningful that in the audio domain. 

Finally, all models generate kicks, snares and cymbals in equal proportions but the generated samples are a bit less diverse than in the original dataset. 

\begin{table*}[]
\begin{tabular}{@{}lcccccccc@{}}
\cmidrule(l){2-9}
                        & \multicolumn{4}{c|}{Test Set}       & \multicolumn{4}{c|}{Noisy Test Set} \\ \cmidrule(l){2-9} 
                        & SDE 400 & ODE 400 & Scipy & DDIM 50 & SDE 400 & ODE 400 & Scipy & DDIM 50 \\ \midrule
VP exp schedule (as in \cite{song2021scorebased}) & 4.11 & 3.96 & 4.54 & 5.11 & 3.36 & 3.45 & 3.04 & 2.87 \\ \midrule
VP cos schedule         & 1.29    & 1.10    & 2.82  & 1.56    & 1.76    & 2.06    & 1.84  & 1.75    \\ \midrule
sub-VP cos schedule     & 1.34    & $\mathbf{0.98}$    & 3.08  & 3.36    & 1.71    & 1.56    & 1.81  & $\mathbf{1.49}$    \\ \midrule
sub-VP 1-1 cos schedule & 1.41    & 1.23    & 2.92  & 2.93    & 1.67    & 2.45    & 2.11  & 1.53    \\ \midrule
sub-VP 1-2 cos schedule & 1.69    & 1.51    & 1.66  & 5.24    & 2.22    & 2.85    & 1.43  & 3.96    \\ \bottomrule
\end{tabular}
\caption{FAD comparison (lower is better)}
\label{FADs}
\end{table*}

\subsection{Interactive sound design}
Audio samples for all experiments described in this section can be heard on the accompanying website: \url{https://crash-diffusion.github.io/crash/}.
\subsubsection{Interpolations}
The relative lack of diversity of the unconditional generation is not dramatic since the model can still perform interactive sound design by modifying existing samples from the dataset. In order to do that, we apply the forward ODE to an existing sound and obtain its corresponding noise in the latent space of isotropic Gaussians. As presented in Fig.~\ref{fig:interpolation_schema}, we can perform spherical combinations on the latent codes and apply the backward ODE to obtain interpolations. Moreover the reconstructed sounds (at the left and right of the schema) are accurate. 
\begin{figure}[h!]
    \includegraphics[scale=0.2]{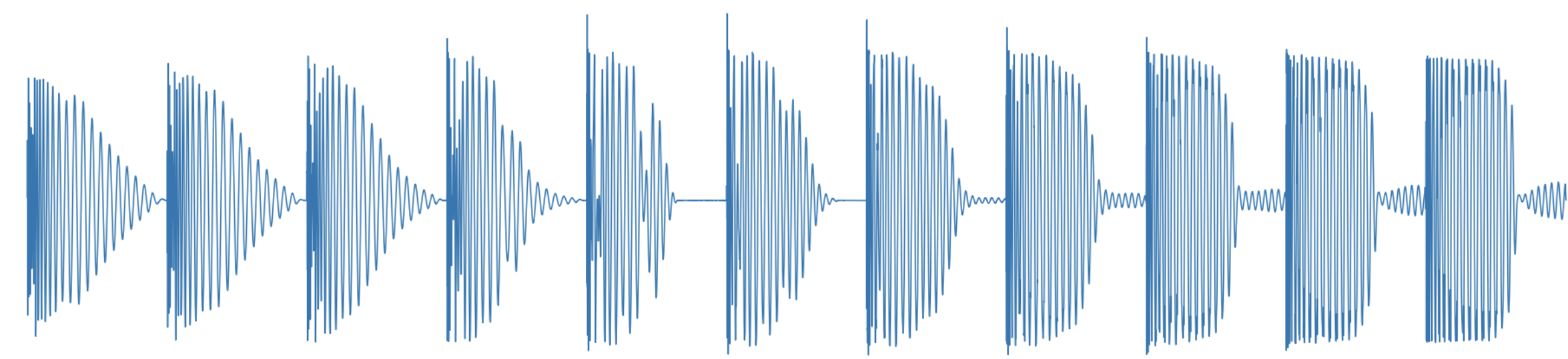}
    \caption{Interpolations between two kicks (at the left and right)}
    \label{fig:interpolations_kick}
\end{figure}

\subsubsection{Obtaining Variations of a Sound by Noising it and Denoising it via SDE}
Let's take a sound $\mathbf{x}(0)$. We can noise it at a desired noise level $\sigma(t)$ via $\mathbf{x}(t) = m(t) \mathbf{x}(0) + \sigma(t) \epsilon$ and then denoise it with a SDE discretization from t to 0. We obtain then variations of the original sound.

\subsubsection{Inpainting}
\begin{figure}[h!]
    \includegraphics[scale=0.35]{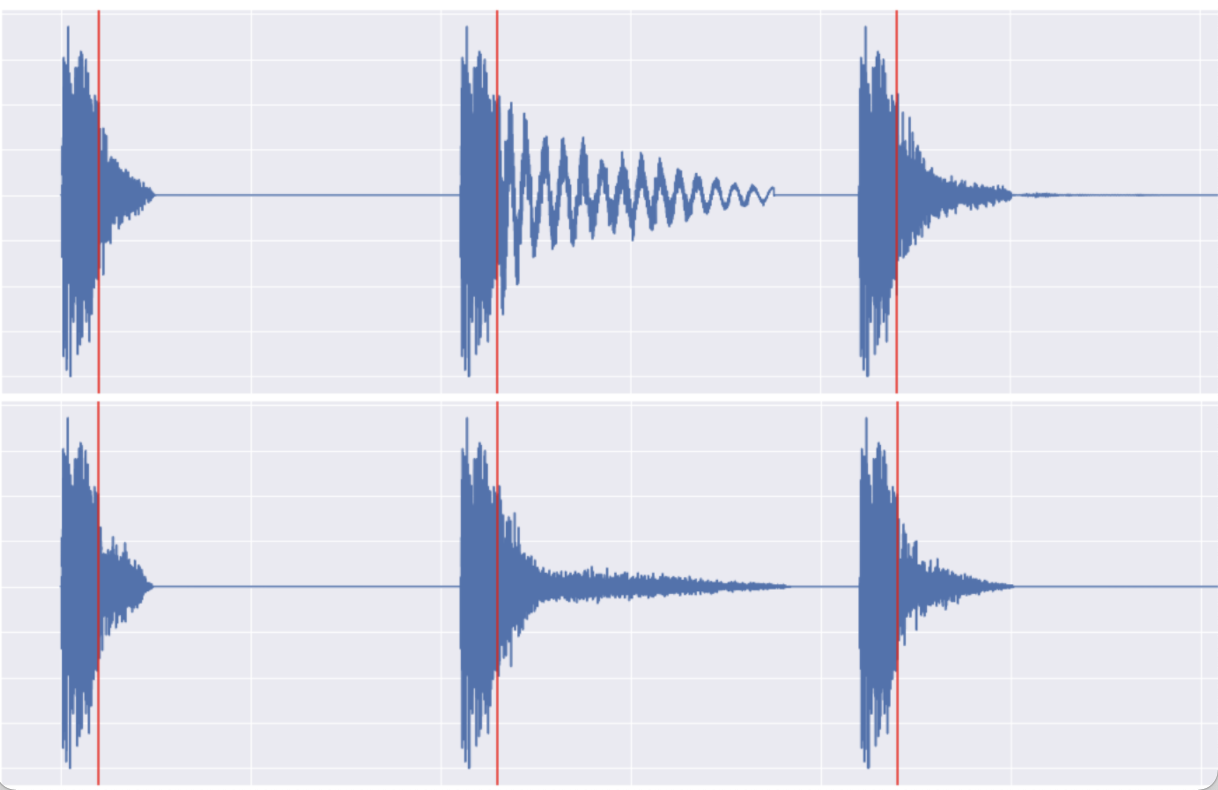}
    \caption{Six Inpaintings on the end of a snare sound}
    \label{fig:inpainting_schema}
\end{figure}

We can also perform inpainting on a sound in order to regenerate any desired part. We show this method on Fig.~\ref{fig:inpainting_schema} where we regenerate 6 endings of a snare sound. 

This provides an innovative way to generate a variety of plausible sounds starting with the same attack.
\subsubsection{Class-Conditioning and Class-Mixing with a Classifier}
We trained a noise-conditioned classifier on the 3 classes (kick, snare, cymbal) and used it to generate class-conditioned and class-mixing generation. Once again, by using the latent representation of a sound we can regenerate it (via ODE) with control on its "kickiness, snariness or cymbaliness". 

\begin{figure}[h!]
    \includegraphics[scale=0.22]{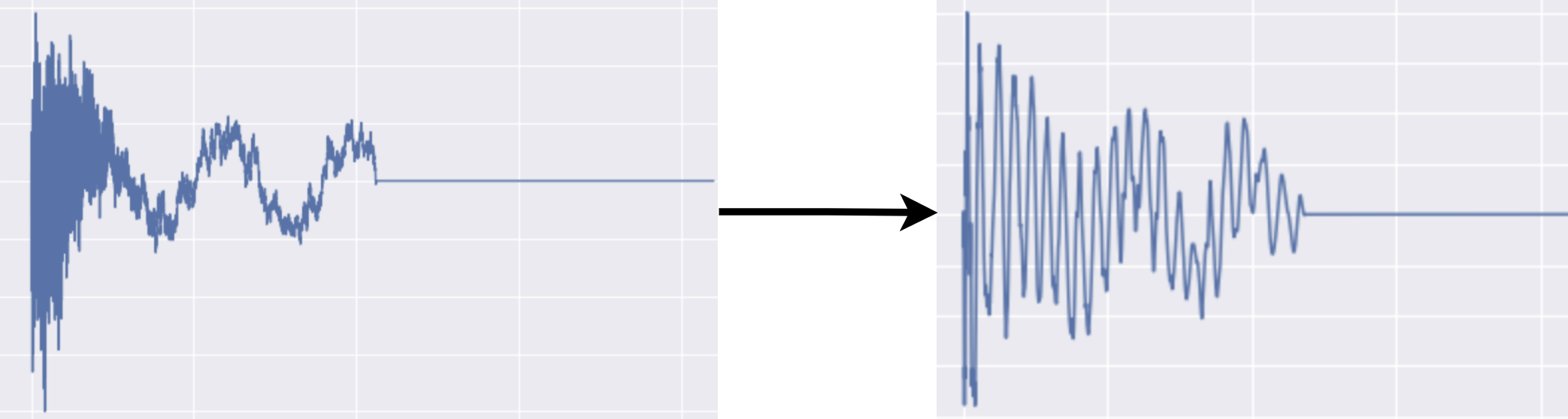}
    \caption{Transformation of a cymbal into a kick via class-conditioning ODE}
\end{figure}

\begin{figure}[h!]
    \includegraphics[scale=0.19]{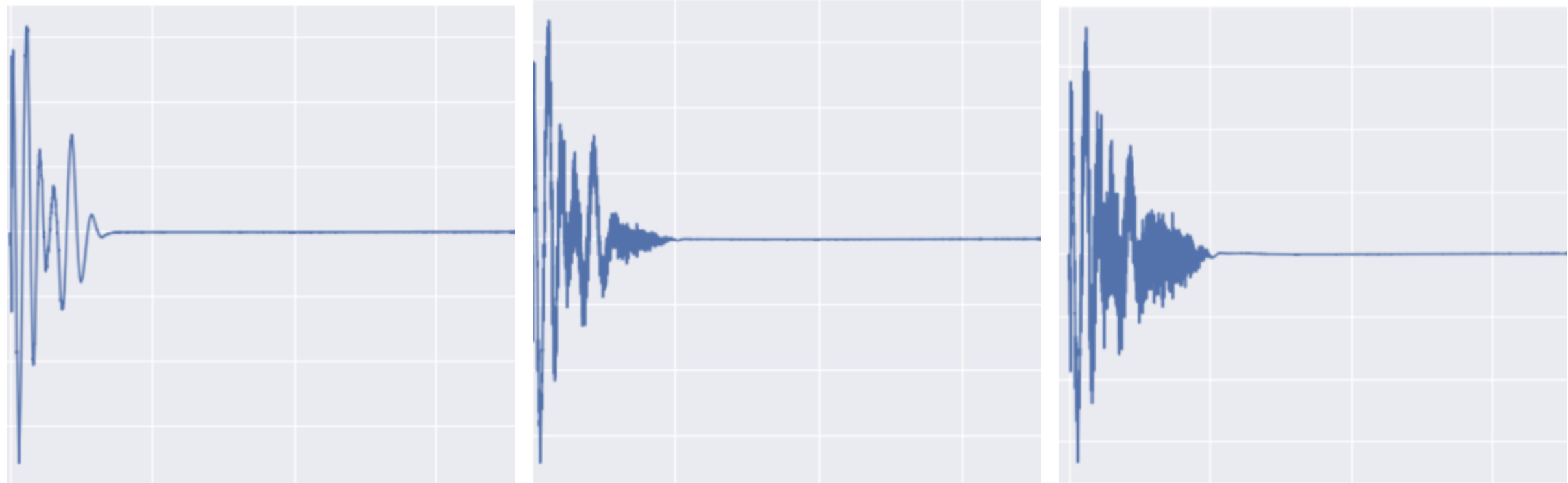}
    \caption{Modifying a kick to make it sounds more "snary" via class-mixing}
    \label{fig:snarification}
\end{figure}

\section{Conclusion}
We presented CRASH, a score-based generative model for the generation of raw audio based on the latest developments in modeling diffusion processes via SDEs. We proposed novel SDEs, well-suited to drum sound generation with high-resolution, together with an efficient architecture for estimating the score function. We showcased how the many controllable sampling schemes offered new perspectives for interactive sound design. In particular, our proposed \emph{class-mixing} strategy allows the controllable creation of convincing "hybrid" sounds that would be hard to obtain with conventional means. We hope that these new methods will contribute to enrich the workflow of music producers. 

\bibliographystyle{unsrtnat}
\bibliography{ms}

\begin{thebibliography}{32}
\providecommand{\natexlab}[1]{#1}
\providecommand{\url}[1]{\texttt{#1}}
\expandafter\ifx\csname urlstyle\endcsname\relax
  \providecommand{\doi}[1]{doi: #1}\else
  \providecommand{\doi}{doi: \begingroup \urlstyle{rm}\Url}\fi

\bibitem[Vasquez and Lewis(2019)]{vasquez2019melnet}
Sean Vasquez and Mike Lewis.
\newblock Melnet: A generative model for audio in the frequency domain, 2019.

\bibitem[Engel et~al.(2019)Engel, Agrawal, Chen, Gulrajani, Donahue, and
  Roberts]{engel2019gansynth}
Jesse Engel, Kumar~Krishna Agrawal, Shuo Chen, Ishaan Gulrajani, Chris Donahue,
  and Adam Roberts.
\newblock Gansynth: Adversarial neural audio synthesis, 2019.

\bibitem[van~den Oord et~al.(2016)van~den Oord, Dieleman, Zen, Simonyan,
  Vinyals, Graves, Kalchbrenner, Senior, and Kavukcuoglu]{oord2016wavenet}
Aaron van~den Oord, Sander Dieleman, Heiga Zen, Karen Simonyan, Oriol Vinyals,
  Alex Graves, Nal Kalchbrenner, Andrew Senior, and Koray Kavukcuoglu.
\newblock Wavenet: A generative model for raw audio, 2016.

\bibitem[Mehri et~al.(2017)Mehri, Kumar, Gulrajani, Kumar, Jain, Sotelo,
  Courville, and Bengio]{mehri2017samplernn}
Soroush Mehri, Kundan Kumar, Ishaan Gulrajani, Rithesh Kumar, Shubham Jain,
  Jose Sotelo, Aaron Courville, and Yoshua Bengio.
\newblock Samplernn: An unconditional end-to-end neural audio generation model,
  2017.

\bibitem[Prenger et~al.(2018)Prenger, Valle, and
  Catanzaro]{prenger2018waveglow}
Ryan Prenger, Rafael Valle, and Bryan Catanzaro.
\newblock Waveglow: A flow-based generative network for speech synthesis, 2018.

\bibitem[Gritsenko et~al.(2020)Gritsenko, Salimans, van~den Berg, Snoek, and
  Kalchbrenner]{gritsenko2020spectral}
Alexey~A. Gritsenko, Tim Salimans, Rianne van~den Berg, Jasper Snoek, and Nal
  Kalchbrenner.
\newblock A spectral energy distance for parallel speech synthesis, 2020.

\bibitem[Donahue et~al.(2019)Donahue, McAuley, and
  Puckette]{donahue2019adversarial}
Chris Donahue, Julian McAuley, and Miller Puckette.
\newblock Adversarial audio synthesis, 2019.

\bibitem[Nistal et~al.(2020)Nistal, Lattner, and Richard]{nistal2020drumgan}
J.~Nistal, S.~Lattner, and G.~Richard.
\newblock Drumgan: Synthesis of drum sounds with timbral feature conditioning
  using generative adversarial networks, 2020.

\bibitem[Aouameur et~al.(2019)Aouameur, Esling, and
  Hadjeres]{aouameur2019neural}
Cyran Aouameur, Philippe Esling, and Ga{\"e}tan Hadjeres.
\newblock Neural drum machine: An interactive system for real-time synthesis of
  drum sounds.
\newblock In \emph{International Conference on Computational Creativity}, 2019.

\bibitem[Bazin et~al.(2021)Bazin, Hadjeres, Esling, and
  Malt]{bazin2021spectrogram}
Th{\'e}is Bazin, Ga{\"e}tan Hadjeres, Philippe Esling, and Mikhail Malt.
\newblock Spectrogram inpainting for interactive generation of instrument
  sounds.
\newblock \emph{arXiv preprint arXiv:2104.07519}, 2021.

\bibitem[Razavi et~al.(2019)Razavi, Oord, and Vinyals]{razavi2019generating}
Ali Razavi, Aaron van~den Oord, and Oriol Vinyals.
\newblock Generating diverse high-fidelity images with vq-vae-2.
\newblock \emph{arXiv preprint arXiv:1906.00446}, 2019.

\bibitem[Vincent(2011)]{vincent2011connection}
Pascal Vincent.
\newblock A connection between score matching and denoising autoencoders, 2011.

\bibitem[Ho et~al.(2020)Ho, Jain, and Abbeel]{ho2020denoising}
Jonathan Ho, Ajay Jain, and Pieter Abbeel.
\newblock Denoising diffusion probabilistic models, 2020.

\bibitem[Song and Ermon(2019)]{song2019generative}
Yang Song and Stefano Ermon.
\newblock Generative modeling by estimating gradients of the data distribution.
\newblock In \emph{Advances in Neural Information Processing Systems}, pages
  11895--11907, 2019.

\bibitem[Song et~al.(2021{\natexlab{a}})Song, Sohl-Dickstein, Kingma, Kumar,
  Ermon, and Poole]{song2021scorebased}
Yang Song, Jascha Sohl-Dickstein, Diederik~P. Kingma, Abhishek Kumar, Stefano
  Ermon, and Ben Poole.
\newblock Score-based generative modeling through stochastic differential
  equations, 2021{\natexlab{a}}.

\bibitem[Song et~al.(2021{\natexlab{b}})Song, Meng, and Ermon]{song2021ddim}
Jiaming Song, Chenlin Meng, and Stefano Ermon.
\newblock Denoising diffusion implicit models.
\newblock In \emph{International Conference on Learning Representations},
  2021{\natexlab{b}}.
\newblock URL \url{https://openreview.net/forum?id=St1giarCHLP}.

\bibitem[Kong et~al.(2021)Kong, Ping, Huang, Zhao, and
  Catanzaro]{kong2021diffwave}
Zhifeng Kong, Wei Ping, Jiaji Huang, Kexin Zhao, and Bryan Catanzaro.
\newblock Diffwave: A versatile diffusion model for audio synthesis, 2021.

\bibitem[Chen et~al.(2020)Chen, Zhang, Zen, Weiss, Norouzi, and
  Chan]{chen2020wavegrad}
Nanxin Chen, Yu~Zhang, Heiga Zen, Ron~J. Weiss, Mohammad Norouzi, and William
  Chan.
\newblock Wavegrad: Estimating gradients for waveform generation, 2020.

\bibitem[Warden(2018)]{warden2018speech}
Pete Warden.
\newblock Speech commands: A dataset for limited-vocabulary speech recognition,
  2018.

\bibitem[Sohl-Dickstein et~al.(2015)Sohl-Dickstein, Weiss, Maheswaranathan, and
  Ganguli]{sohldickstein2015deep}
Jascha Sohl-Dickstein, Eric~A. Weiss, Niru Maheswaranathan, and Surya Ganguli.
\newblock Deep unsupervised learning using nonequilibrium thermodynamics, 2015.

\bibitem[Tov et~al.(2021)Tov, Alaluf, Nitzan, Patashnik, and
  Cohen{-}Or]{encoder2021tov}
Omer Tov, Yuval Alaluf, Yotam Nitzan, Or~Patashnik, and Daniel Cohen{-}Or.
\newblock Designing an encoder for stylegan image manipulation.
\newblock \emph{CoRR}, abs/2102.02766, 2021.
\newblock URL \url{https://arxiv.org/abs/2102.02766}.

\bibitem[Härkönen et~al.(2020)Härkönen, Hertzmann, Lehtinen, and
  Paris]{harkonen2020ganspace}
Erik Härkönen, Aaron Hertzmann, Jaakko Lehtinen, and Sylvain Paris.
\newblock Ganspace: Discovering interpretable gan controls, 2020.

\bibitem[Mirza and Osindero(2014)]{mirza2014conditional}
Mehdi Mirza and Simon Osindero.
\newblock Conditional generative adversarial nets, 2014.

\bibitem[Anderson(1982)]{anderson1982reverse}
Brian D~O Anderson.
\newblock Reverse-time diffusion equation models, 1982.

\bibitem[Durkan and Song(2021)]{durkan2021maximum}
Conor Durkan and Yang Song.
\newblock On maximum likelihood training of score-based generative models,
  2021.

\bibitem[Nichol and Dhariwal(2021)]{nichol2021improved}
Alex Nichol and Prafulla Dhariwal.
\newblock Improved denoising diffusion probabilistic models, 2021.

\bibitem[Meseguer-Brocal and
  Peeters(2019)]{MeseguerBrocal2019ConditionedUNetIA}
Gabriel Meseguer-Brocal and G.~Peeters.
\newblock Conditioned-u-net: Introducing a control mechanism in the u-net for
  multiple source separations.
\newblock \emph{ArXiv}, abs/1907.01277, 2019.

\bibitem[Tancik et~al.(2020)Tancik, Srinivasan, Mildenhall, Fridovich-Keil,
  Raghavan, Singhal, Ramamoorthi, Barron, and Ng]{tancik2020fourier}
Matthew Tancik, Pratul~P. Srinivasan, Ben Mildenhall, Sara Fridovich-Keil,
  Nithin Raghavan, Utkarsh Singhal, Ravi Ramamoorthi, Jonathan~T. Barron, and
  Ren Ng.
\newblock Fourier features let networks learn high frequency functions in low
  dimensional domains, 2020.

\bibitem[Perez et~al.(2017)Perez, Strub, de~Vries, Dumoulin, and
  Courville]{perez2017film}
Ethan Perez, Florian Strub, Harm de~Vries, Vincent Dumoulin, and Aaron
  Courville.
\newblock Film: Visual reasoning with a general conditioning layer, 2017.

\bibitem[Ronneberger et~al.(2015)Ronneberger, Fischer, and
  Brox]{ronneberger2015unet}
Olaf Ronneberger, Philipp Fischer, and Thomas Brox.
\newblock U-net: Convolutional networks for biomedical image segmentation,
  2015.

\bibitem[Song and Ermon(2020)]{song2020improved}
Yang Song and Stefano Ermon.
\newblock Improved techniques for training score-based generative models, 2020.

\bibitem[Kilgour et~al.(2019)Kilgour, Zuluaga, Roblek, and
  Sharifi]{kilgour2019frechet}
Kevin Kilgour, Mauricio Zuluaga, Dominik Roblek, and Matthew Sharifi.
\newblock Fr\'echet audio distance: A metric for evaluating music enhancement
  algorithms, 2019.

\end{thebibliography}
\newpage
\appendix

\section{Proof of the reparametrization}
\label{sec:reparam}
\subsection{Forward SDE}
By dividing Eq.~\ref{eq:sde1} by $m$ we have
\begin{equation}
        \frac{\dd{\mathbf{x}}}{m(t)} = \frac{f(t)}{m(t)} \mathbf{x} \dd{t} + \frac{g(t)}{m(t)} \dd{\mathbf{w}},
\label{eq:sde_div_by_m}
\end{equation}
moreover, the first equation of Eq.~\ref{eq:system_m_sigma} gives
\begin{equation}
    f(t) \dd{t} = \frac{\dd{m}}{m(t)}
\end{equation}
and Eq.~\ref{eq:sde_div_by_m} becomes
\begin{equation}
    \frac{m(t) \dd{\mathbf{x}}}{m^2(t)} = \frac{\mathbf{x} \dd{m}}{m^2(t)} + \frac{g(t)}{m(t)} \dd{\mathbf{w}}.
\label{eq:sde_transformed}
\end{equation}

By dividing the second relation of Eq.~\ref{eq:system_m_sigma} by $m^2(t)$, we obtain the equation:
\begin{equation}
    \frac{2 \sigma(t) \sigma'(t)}{m^2(t)} = 2 f(t) \frac{\sigma^2(t)}{m^2(t)}+ \frac{g^2(t)}{m^2(t)},
\end{equation}
since $f(t) = \frac{m'(t)}{m(t)}$, we have then:
\begin{equation}
    \frac{2 \sigma(t)\sigma'(t)m^2(t) - 2 m'(t)m(t)\sigma^2(t)}{m^4(t)} = \frac{g^2(t)}{m^2(t)},
\end{equation}
and since $g$ and $m$ are positive, this can be rewritten into:
\begin{equation}
    \frac{g(t)}{m(t)} = \sqrt{\dv{t}(\frac{\sigma^2(t)}{m^2(t)})}.
\label{eq:g_over_m}
\end{equation}
Equation \ref{eq:sde_transformed} becomes then:
\begin{equation}
    \dd{\left(\frac{\mathbf{x}}{m}\right)} = \sqrt{\dv{t}(\frac{\sigma^2}{m^2})} \dd{\mathbf{w}}.
\end{equation}

\subsection{Reverse time SDE}

According to Eq.~\ref{eq:reverse_sde}, the associated reverse time SDE is:

\begin{equation}
    \dd{\left(\frac{\mathbf{x}}{m}\right)}=-\dv{t}(\frac{\sigma^2}{m^2}) \nabla_{\frac{\mathbf{x}}{m}} \log q_t \left(\frac{\mathbf{x}}{m} \right) \dd{t}+ \sqrt{\dv{t}(\frac{\sigma^2}{m^2})} \dd{\mathbf{\Tilde{w}}}
\end{equation}
where     
$q_t(\frac{\mathbf{x}(t)}{m(t)} \mid \frac{\mathbf{x}(0)}{m(0)}) = \mathcal{N}(\frac{\mathbf{x}(t)}{m(t)}; \mathbf{x}(0), \frac{\sigma^2(t)}{m^2(t)}\mathbf{I})$.

A quick calculus shows that: 
\begin{equation}
    \nabla_{\frac{\mathbf{x}}{m}} \log q_t \left(\frac{\mathbf{x(t)}}{m(t)} \mid \frac{\mathbf{x}(0)}{m(0)}\right) = m(t) \nabla_{\mathbf{x}} \log p_{t}(\mathbf{x}(t) \mid \mathbf{x}(0)).
\end{equation}


Using the fact that $m(0)=1$, by integrating over $p_\text{data}$ we obtain:
\begin{equation}
    \nabla_{\frac{\mathbf{x}}{m}} \log q_t \left(\frac{\mathbf{x}}{m} \right) = m(t) \nabla_{\mathbf{x}} \log p_{t}(\mathbf{x}(t)).
\end{equation}

Then, by writing $\mathbf{\epsilon}(\mathbf{x}, \sigma) := - \sigma(t)\nabla_{\mathbf{x}} \log p_t(\mathbf{x})$ we obtain:
\begin{equation}
\begin{aligned}
    -\dv{t}(\frac{\sigma^2}{m^2}) \nabla_{\frac{\mathbf{x}}{m}} \log q_t \left(\frac{\mathbf{x}}{m} \right) = \dv{t}(\frac{\sigma^2}{m^2})  \frac{m}{\sigma} \mathbf{\epsilon}(\mathbf{x}, \sigma) \\
    = 2 \dv{t}(\frac{\sigma}{m}) \mathbf{\epsilon}(\mathbf{x}, \sigma)
\end{aligned}
\end{equation}
which gives us the reverse-time SDE:

\begin{equation}
    \dd{\left(\frac{\mathbf{x}}{m}\right)}= 2 \dv{t}(\frac{\sigma}{m})\mathbf{\epsilon}(\mathbf{x}, \sigma) \dd{t}+ \sqrt{\dv{t}(\frac{\sigma^2}{m^2})} \dd{\mathbf{\Tilde{w}}}.
\end{equation}

\subsection{Associated ODE}
Finally, the associated ODE comes directly with a factor $\frac{1}{2}$ behind the drift of the reverse-time SDE:

\begin{equation}
    \dd{\left(\frac{\mathbf{x}}{m}\right)}=  \dv{t}(\frac{\sigma}{m})\mathbf{\epsilon}(\mathbf{x}, \sigma) \dd{t}.
\end{equation}

\section{About the link between an affine $\log$ SNR and the Maximum Likelihood Estimation}
Let's consider a SNR such as $\log \text{SNR}(t)$ is affine. Then, we have $\frac{\sigma(t)^2}{m(t)^2} = e^{at+b}$. By using Eq.~\ref{eq:g_over_m}, we obtain $g(t)^2 = m(t)^2 a e^{at+b} = a m(t)^2 \frac{\sigma(t)^2}{m(t)^2} = a \sigma(t)^2$. 
This means that having an affine $\log \text{SNR}(t)$ is a situation where the maximum likelihood estimator is equivalent to the ancestral training method (where $\lambda(t) \propto \sigma(t)^2$).

\section{Architectural details}
\label{sec:details}
\subsection{Random Fourier Features Embeddings and FiLM}
First, the embedding of $\sigma$ is done with Random Fourier Features (RFF):
\begin{equation}
[\cos(2\pi f_1 \sigma), …, \cos(2\pi f_N \sigma), \sin(2\pi f_1 \sigma), …, \sin(2\pi f_N \sigma)]
\end{equation} with $N = 32$ and $f_1, …, f_N$ are sampled from a one-dimensional Gaussian with zero mean and variance 16 and fixed for the whole training. Unlike other architectures, we make the choice of encoding $\sigma$ instead of $t$ because the network does not have to learn the function $\sigma(t)$. The RFF are followed by a MLP. For each DBlock, the embedding of $\sigma$ is fed into a Linear layer of output size $2C_\text{out}$ that is chunked in order to obtain $\gamma(\sigma)$ and $\beta(\sigma)$ that are the factors of the FiLM-like operation.

\begin{figure}[h!]
    \centering
    \includegraphics[scale=0.28]{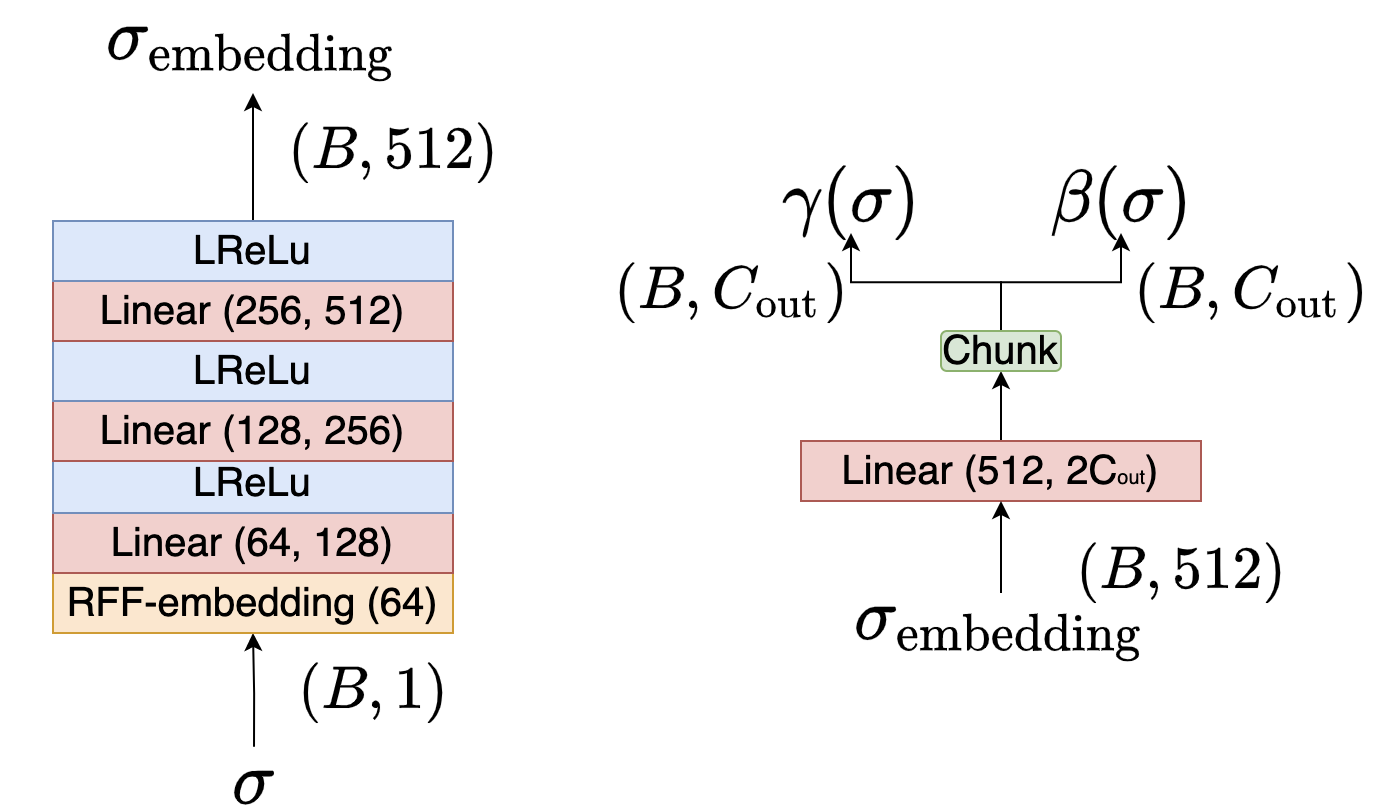}
    \caption{A Downsampling Block}
    \label{fig:my_label}
\end{figure}

\subsection{Downsampling and Upsampling Blocks}
The DBlocks begin with a Downsampling task: for instance if the downsampling factor is 3, only the first element of each sequence of 3 is kept. Then there is a first 1D-Convolution with an output channel size $C_\text{out}$ equal to the output channel size of the Block. Then, a FiLM-like \cite{perez2017film} operation is done to the signal $x$ of shape $(B, C_\text{out}, \frac{T}{f})$: 
\begin{equation}
    \gamma(\sigma) \odot x + \beta(\sigma)
\end{equation}
After that, the signal goes into three 1-D convolutions. For the four 1-D convolutions of the main Block the dilation factors are $1, 2, 4, 8$ and the kernel size is $3$. For the residual Block the convolution has an output channel size $C_\text{out}$ and a kernel size of $1$.

\begin{figure}[h!]
    \centering
    \includegraphics[scale=0.28]{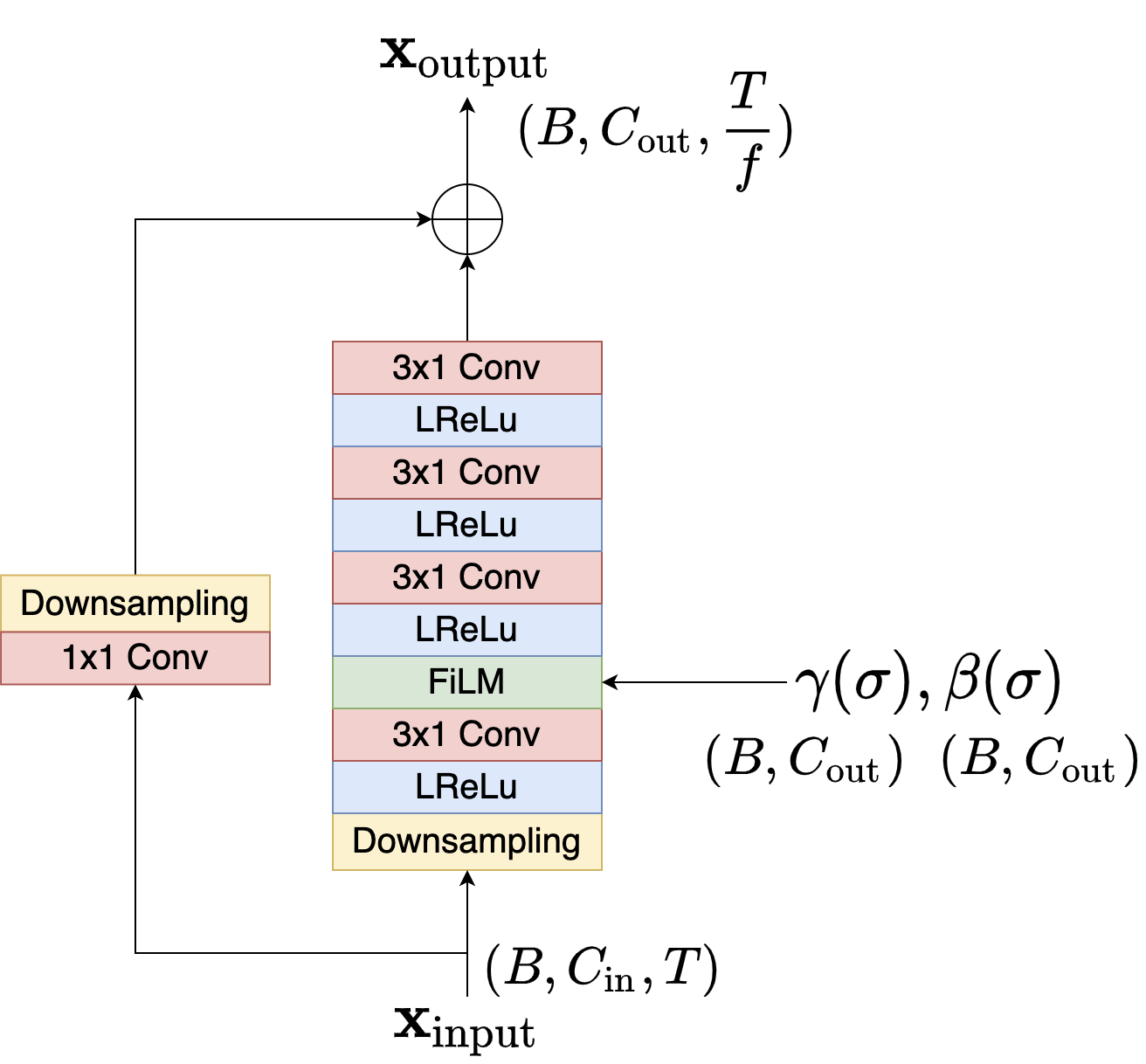}
    \caption{A Downsampling Block}
    \label{fig:my_label}
\end{figure}

The UBlocks are similar to the DBlocks but there are some little differences. The input $\mathbf{x}_\text{input}$ is concatenated with the output of the DBlock of the same level $\mathbf{x}_\text{DBlock}$. The concatenation which has the shape $(B, 2C_\text{in}, T)$ is upsampled (by repeating each number $f$ time) and then goes into a 1-D convolution with dilation $1$ number of output channels $C_\text{out}$. The following convolutions have dilation factors of $2, 4, 8$.

\begin{figure}[h!]
    \centering
    \includegraphics[scale=0.28]{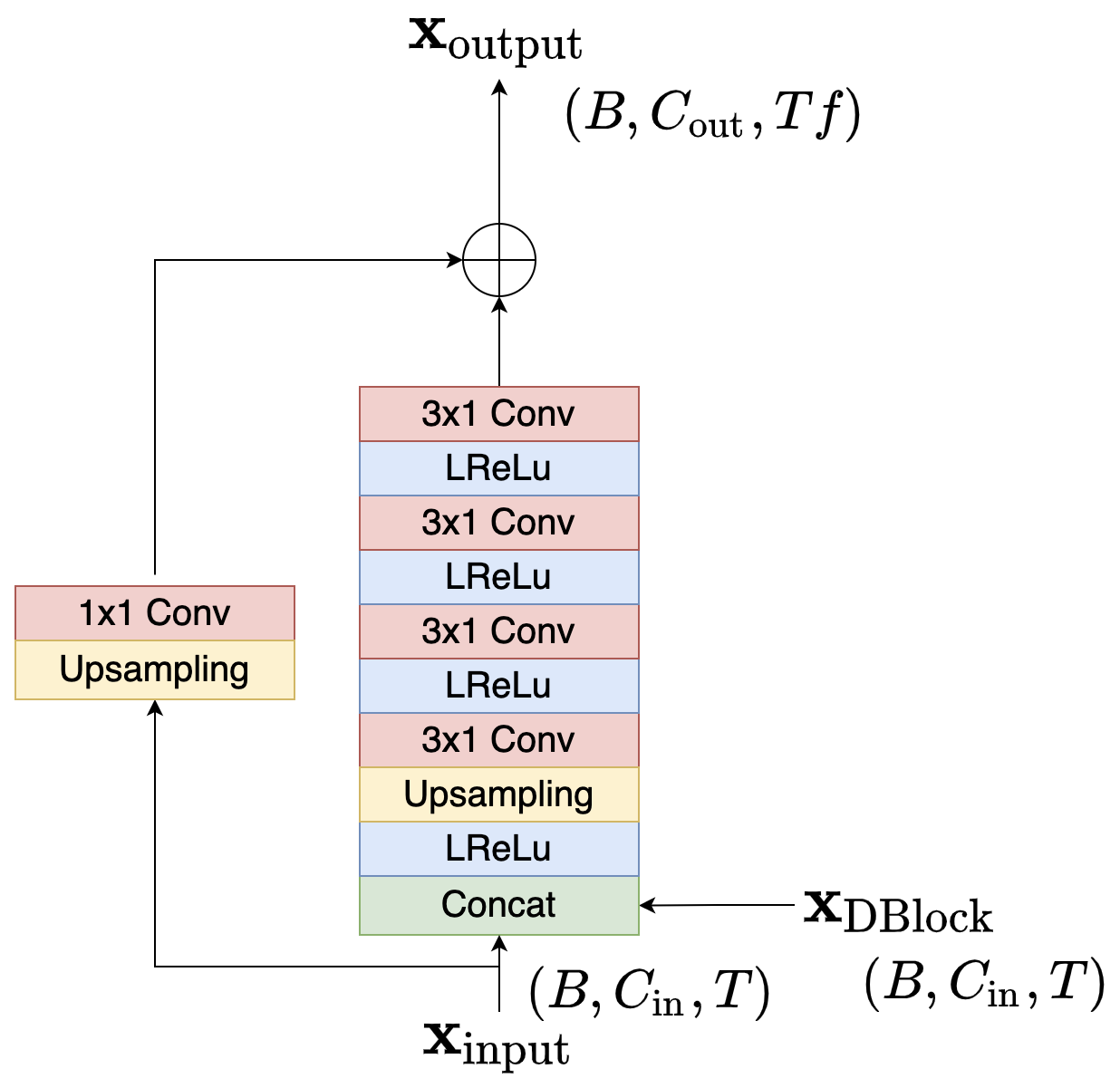}
    \caption{An Upsampling Block}
    \label{fig:my_label}
\end{figure}

\subsection{About the U-Net}
The downsampling and upsampling factors of our U-Net are $2, 2, 3, 5, 5$ which means that the input audio must have a length divisible by $300$. The number of channels of the outputs of the DBlocks are: $128, 128, 256, 512, 512$.

\subsection{About the noise-conditioned classifier}
\label{sec:ncclassif}
The noise-conditioned classifier is constituted of one 1D convolution with 32 channels with a kernel size of 5 and a padding of 2. Followed by 5 DBlocks with downsampling factors of $4, 3, 5, 25, 14$ with outputs channels of size $128, 256, 512, 512, 512$. There is a final linear layer of output size $3$ followed by a softmax. 

\section{About the generalization of sub-VP}
\label{sec:sub-vp_gen}
Considering the following forward SDE : 
\begin{equation}
    \dd{\mathbf{x}} = -\frac{1}{2}\beta(t)\mathbf{x}\dd{t}+ g(t) \dd{\mathbf{w}}
\end{equation}

The associated following system for $m$ and $\sigma$ is :
\begin{equation}
    \left\{
        \begin{array}{ll}
            \dv{m}{t} = -\frac{1}{2}\beta(t) m(t) \\
            \dv{\sigma^2(t)}{t} = -\beta(t) \sigma^2(t) + g^2(t) 
        \end{array}
    \right.
\label{system_m_sigma}
\end{equation}

with the following conditions :
\begin{equation}
    \left\{
        \begin{array}{ll}
            m(0) = 1 \\
            \sigma^2(0) = 0
        \end{array}
    \right.
\end{equation}

The solutions are :
\begin{equation}
    \left\{
        \begin{array}{ll}
            m(t) = e^{-\int_{0}^{t} \frac{1}{2}\beta(s) \dd{s}} \\
            \sigma^2(t) = e^{- \int_{0}^{t} \beta(s) \dd{s}} \int_{0}^{t} g^2(u) e^{\int_{0}^{u} \beta(s) \dd{s}} \dd{u}
        \end{array}
    \right.
\label{solution_m_sigma2}
\end{equation}

Now, by applying the method described in \ref{sec:discussion}, if we have chosen $\sigma(t)$ and a relation between $m$ and $\sigma$ in the form 
\begin{equation}
    m = (1 - \sigma^\gamma)^\eta
\end{equation}
In order to calculate $\beta(t)$ and $g(t)$, the first equation of Eq.~\ref{system_m_sigma} gives the expression of $\beta$ function of $\sigma$ and its derivative $\sigma'$, the second equation gives $g(t)$:
\begin{equation}
    \beta(t) = \frac{2\eta \gamma \sigma'(t) \sigma^{\gamma - 1}(t)}{1-\sigma^\gamma(t)} \
\end{equation}
\begin{equation}
g(t) = \sqrt{2 \sigma^{'}(t) \sigma(t)(\frac{\gamma \eta \sigma^{\gamma}(t)}{1 - \sigma^\gamma(t)} +1)}
\end{equation}
and we can rewrite it as:
\begin{equation}
\begin{split}
    g(t) = (\beta(t)[(1 - e^{-\frac{1}{2\eta}\int_{0}^{t} \beta(s) \dd{s}})^\frac{2}{\gamma} \\
    + \frac{1}{\gamma \eta} e^{-\frac{1}{2\eta}\int_{0}^{t} \beta(s) \dd{s}} (1 - e^{-\frac{1}{2\eta}\int_{0}^{t} \beta(s) \dd{s}})^{\frac{2}{\gamma} - 1}])^{\frac{1}{2}}
\end{split}
\end{equation}

\end{document}